# Analysis of central Hox protein types across bilaterian clades: On the origin of central Hox proteins from an Antennapedia/Hox7-like ancestor

Hueber, Stefanie D.[1,3]; Djordjevic, Michael A.[1]; Gunter, Helen[2]; Rauch, Jens[3]; Weiller, Georg F.[1]; Frickey, Tancred[1,3,*]

[1] ARC Centre of Excellence (CILR), Research School of Biology, College of Medicine, Biology and the Environment, RN Robertson Building (Bldg 46) Biology Place, The Australian National University, Acton, Canberra ACT 0200, Australia.
[2] Lehrstuhl fuer Zoologie und Evolutionsbiologie, Dept. of Biology, Universität Konstanz, D-78457 Konstanz, Germany.
[3] Applied Bioinformatics, Dept. of Biology, Universität Konstanz, D-78457 Konstanz, Germany.
[*] Corresponding author

Email addresses:
   SDH: stefanie.hueber@uni-konstanz.de
   MAD: michael.djordjevic@anu.edu.au
   HG: helen.gunter@uni-konstanz.de
   JR: jens.rauch@uni-konstanz.de
   GFW: georg.weiller@anu.edu.au
   TF: tancred.frickey@uni-konstanz.de

**Key words:** Hox protein, Antennapedia (Antp), Hox7, evolution, classification, functional equivalence, Hox protein types, echinoderm/hemichordate Hox6, echinoderm/hemichordate Hox7

**Abstract:**

Background: Hox proteins are one of the best studied sets of transcription factors in developmental biology. They are the major determinants for establishing morphological differences along the anterior-posterior axis of animals and are generally regarded as highly conserved in function. This view is mostly based on experiments comparing a few (anterior) Hox proteins, however, the extent to which central or abdominal Hox proteins share conserved sequence features or functions remains largely unexplored.

Results: To shed light on the origin and functional divergence of the central Hox proteins, we combine a powerful bioinformatics tool (CLANS) with a large-scale phylogeny of species. CLANS is used to differentiate between the various types of central Hox protein sequences present in different species, while the phylogeny provides an evolutionary context to the analysis. The combination of both enables us to infer the relative timepoint at which a given type of central Hox proteins arose. We identify seven distinct central Hox protein sequence types, only one of which is common to all protostome and deuterostome clades (Antp/Hox7).

Conclusion: Together, these results lead us to suggest a reevaluation of the usefulness of the increasingly depicted synteny-based classification scheme that assumes a one-to-one orthology between protostome and deuterostome central Hox proteins. Instead, we propose that the use of sequence-based classification schemes capable of resolving the central and posterior Hox proteins provides a more promising and biologically meaningful alternative to resolving these groups.

This analysis, which provides a unique overview of the Hox protein sequence types present across protostomes and deuterostomes as well as a relative dating for the emergence of the various central Hox protein types, provides a crucial first step to help shed light on how and when the distinct developmental blueprints for organisms evolved within the evolutionarily immensely successful bilaterian lineage.

**Background**

One of the most exciting puzzles in developmental biology is provided by the question of how the set of Hox protein transcription factors induce their respective highly visible and distinct morphologies along the anterior-posterior axis of bilaterian organisms [1]. Each Hox protein, encoded by a gene in a given Hox gene cluster, differs in its amino-acid sequence and its ability to regulate downstream genes [2, 3] and many Hox protein sequences show a greater similarity to proteins encoded by genes present in the Hox gene clusters of distantly related species than to adjacent genes in the same cluster. Combined with the observation that some Hox proteins from distantly related species could be shown to carry out nearly identical functions (i.e. they were able to induce the same morphological structures and/or regulate the same downstream genes in a given model organism) [4–6], this led to the hypothesis that the last common ancestor of protostomes and deuterostomes must have already possessed a set of multiple differentiated Hox proteins whose functions remained mostly conserved [2]. To better understand the link between the similarities in sequence and the respective similarities in function of the proteins, the available classifications and functional comparisons performed for this protein family are often used as a basis to: 1) predict which of the Hox proteins carry out similar functions in the organisms being compared 2) identify sequence elements conserved across or specific to certain types of Hox proteins and 3) identify which of these sequence elements are likely to be responsible for a given function shared by a subset of Hox proteins.

Despite intensive research and the accumulation of vast amounts of sequence and experimental data, technical difficulties have prevented detailed characterization of the evolutionary and functional relationships for the complete set of Hox proteins [7]. For the more anterior Hox proteins (Hox 1-5 in vertebrates) the evolutionary and functional relationships are largely undisputed (see Figure 1, Top) as these assignments are based on numerous sequence analyses [8–11] and functional comparison studies [4–6]. However, even when taking into account the large body of work available, the classification of the central and posterior Hox protein sequences remains ambiguous [11, 12]. We would also like to point out that the definition of what constitutes "central" or "middle" Hox proteins used to vary widely (compare [13] with [14] and [9]) and that current definitions classify the vertebrate Hox4-8 proteins as "central". The assignment of Hox4 group proteins as most similar to Deformed (Dfd) as well as Hox5 group proteins as most similar to Sex combs reduced (Scr), is predominantly accepted (see Figure 1, Top). As the aim of this manuscript is to resolve the sequence-relationships of the central Hox proteins whose assignments are in dispute (see Figure 1, Bottom), we will, in this manuscript, use the original definition of central Hox proteins encompassing only Hox6-8 in vertebrates and Antennapedia (Antp), Ultrabithorax (Ubx) and Abdominal-A (Abd-A) in arthropods [9–11, 15–17].

The most exhaustive sequence-based classification for Hox proteins was published by deRosa et al., in 1999 [9]. This publication provided a set of clear and undisputed assignments for which Hox protein types are present in the protostome clades and these assignments have been supported by subsequent analyses [10, 18, 19]. However, no sequence-based analysis has yet clearly identified or been able to differentiate between all types of central-group Hox proteins present across the protostome and deuterostome lineage as a whole.

By taking into account two additional sequence regions relevant to Hox protein function (see supplementary Figure S1), the YPWM motif and 'linker' region, a previous analysis of ours provided a protein-sequence-data based classification of this protein family [11]. The most crucial aspect of this work was that it provided a classification assigning, into separate groups, the phylogenetically unresolved central-type Hox proteins known to have different developmental functions. However, this work specifically focused on a restricted set of model organisms and species (*Mus musculus*, *Danio rerio*, *Branchiostoma floridae*, *Caenorhabditis elegans* and *Drosophila melanogaster*). As a direct consequence, this previous analysis did not provide the breadth of sampling required to accurately determine how widely the different identified types of Hox protein sequences are represented across the protostome and deuterostome lineages.

In this manuscript, we extend upon our previous work [11] by providing an analysis taking into account all central Hox protein sequences from all species available in the public databases and comparing the distribution of species across the different identified types of central Hox protein sequences. To provide an evolutionary context to the results, we combine the species-distribution information gained in the step above with a recent phylogenetic reconstruction of the major protostome and deuterostome lineages [20]. This approach has three major advantages: 1) The CLANS program [21], employed to identify the different central Hox sequence types present in the protostome and deuterostome lineages, bases its analysis on pairwise sequence alignment data and thereby avoids the errors induced by the difficulty of generating high-quality multiple sequence alignments for sequences that are difficult to align due to varying lengths and compositions. 2) As a further advantage, the CLANS program can work with large numbers of sequences, enabling, for example, the inclusion of all Hox and Hox related sequences for all species for which data is available in the NCBI non-redundant protein database "nr". 3). Combining the information about which species contain which types of central Hox proteins with a phylogeny of the species, enables us to infer the earliest branch point in the cladogram of protostome and deuterostome organisms from which on subsequent lineages contain a given Hox protein type, providing an approximate timing for when it arose.

The central Hox proteins from all species available in the public database 'nr' could be classified into seven distinct sequence similarity groups: one combining species from the protostome and deuterostome lineages (Antp/Hox7), one present in all protostomes but no deuterostomes (Abd-A), one specific to arthropods (Ubx), two specific to vertebrates (Hox6, Hox8) and two previously undescribed sequence similarity groups specific to the echinoderm and hemichordate lineages (Echi/Hemi7, Echi/Hemi8).

**Results**

Figure 2 provides a CLANS map visualizing the all-against-all pairwise sequence similarities of the Hox protein sequences retrieved from the NCBI non-redundant protein database "nr". The left side provides an overview of the similarities across the entire family of Hox proteins, while the right side depicts a more detailed representation of the cluster combining proteins of the Hox4-8, Dfd, Scr, Antp, Abd-A and Ubx sequence types. Sequences are depicted as dots (2629 (left side) and 1095 (right side)) and lines connecting the dots represent the corresponding pairwise sequence similarities. The more similar two sequences are to each other, the darker the line connecting them and the more closely they will be located to each other in the map.

The NCBI GenBank non-redundant protein database 'nr' unfortunately also contains numerous partial or fragment sequences that do not span the entire sequence region upon which this analysis is based (supplementary Figure S1: YPWM-motif, linker region and homeodomain). For example, sequences that lack the YPWM+linker region, are missing one of the most informative regions for classifying central Hox proteins across deuterostome and protostome clades. Consequently, these partial sequences cannot be assigned to a specific Hox type with the same confidence as sequences containing the full YPWM-linker-homeodomain region. We therefore manually identified the sequences containing the complete region of interest (YPWM+linker+Homeodomain) and highlighted these with large colored dots in Figure 2. Partial or fragment sequences are represented by smaller colored dots. Sequences we could not assign to a specific sequence similarity group were left uncolored.

Based on visual inspection, 10 sequence-similarity groups are identifiable in the resulting map (Figure 2): two groups consisting of Hox protein types defined as "anterior" by de Rosa (see Figure 1) [9, 10]: Dfd/Hox4 (light red) and Scr/Hox5 (brown), a group combining *Drosophila* Ftz-like (grey) proteins (often regarded as recently derived from a central Hox protein ancestor and therefore included in this analysis [22–24]), and seven distinct sequence similarity groups for the central Hox proteins. Five of the seven groups encompass the vertebrate and *Drosophila spp.* central-group Hox proteins (Abd-A (green), Ubx (blue), Hox6 (orange), Hox8 (dark red) and Antp+Hox7 (yellow)) [11]. The remaining two groups have not previously been described and consist of central Hox protein types found only in the echinoderm/hemichordate lineages (Echi/Hemi7 colored in beige and Echi/Hemi8 in purple). To identify the overlap between the pairwise-similarity-based grouping of Hox protein sequences and the taxonomic classification of the corresponding species, we identified which species were present in the respective sequence similarity groups. This information was then used as the basis to map the presence of the corresponding central Hox sequence types onto a cladogram depicting the phylogenetic relationships of the major protostome and deuterostome lineages (Figure 3). Based on the comparison of the phylogeny of species with the distribution of species across the identified central Hox sequence types, we observe the following:

**Four central Hox protein types are specific to deuterostomes**

Three separate types of central Hox proteins were recognizable in all examined vertebrate groups, including "basal" vertebrates such as the lamprey *Petromyzon marinus* or the horn shark *Heterodontus francisci* as well as in "higher" vertebrates such as the mouse *Mus musculus*, chicken *Gallus gallus* and zebrafish *Danio rerio*: Hox6 (orange), Hox7 (yellow) and Hox8 (dark red) (Figure 2 & supplementary Figure S2). Sequences grouping with the vertebrate Hox6 and Hox8 proteins could not be detected in any non-vertebrate species, not even in closely branching deuterostomes such as the cephalochordate *Branchiostoma lanceolatum* and the urochordate *Ciona intestinalis* or the more distantly branching hemichordates *Saccoglossus kowalevskii* or *Balanoglossus simodensis* as well as the echinoderms *Strongylocentrotus purpuratus* or *Metacrinus rotundus* (Figure 3 & supplementary Figure S2). All three central Hox proteins from *Branchiostoma lanceolatum* are most similar in sequence to the Antp/Hox7 group (yellow) and, as such, cannot be regarded as either of the Hox6 or Hox8 sequence type. *Ciona intestinalis* contains a single central-type Hox protein and this sequence is most similar to the Antp/Hox7-like sequences. Hemichordates and echinoderms possess two previously undescribed types of central Hox proteins that are specific to their lineage and form two separate sequence-similarity groups: a group that lies peripheral to, but remains most similar in sequence to the Antp/Hox7 group (Echi/Hemi7) (purple) and a second more derived version of an Antp/Hox7 type sequence (Echi/Hemi8) (beige). A third central-type Hox protein present in hemichordates remains clearly identifiable as an Antp/Hox7 type sequence.

In summary, four central Hox protein types are specific to deuterostomes: Hox6 and Hox8 type proteins are only found in the vertebrate lineage while Echi/Hemi7 and Echi/Hemi8 type proteins are only found in the echinoderm and hemichordate lineages.

**Two central Hox protein types are specific to protostomes**

In the arthropod clade, we could assign central Hox protein sequences from insects (*Drosophila melanogaster*, *Tribolium castaneum*), chelicerates (*Cupiennus salei*, *Parasteatoda tepidariorum*, *Ixodes scapularis*) and crustaceans (*Procambarus clarkii*, *Daphnia magna*) to each of three sequence groups: Antp/Hox7-like (yellow), Abd-A-like (green) and Ubx-like (blue) (Figure 2 and supplementary Figure S2). While only a single Antp-like protein could be identified per species in the crustacean lineage, the species in the chelicerate and insect lineages contain two Antp-like proteins. In the chelicerates both proteins are highly similar to Antp type proteins while in the insect lineage one of the proteins is clearly recognizable as an Antp type protein, referred to as *Drosophila* Antp, while the second diverged in sequence and function to a greater extent and is referred to as *Drosophila* Ftz. This Ftz protein (grey) was previously shown to no longer have a Hox-like expression and function in Drosophila embryos (Ftz is a pair-rule protein previously predicted to have been derived from a Hox protein ancestor [22–24]).

In each of the three well-described nematode model organisms, *Caenorhabditis elegans*, *Caenorhabditis briggsae* and *Pristionchus pacificus*, we could identify only a single, further unclassifiable, central-like Hox protein (MAB-5 protein type). The lophotrochozoa, represented in Figure 3 by the annelid *Capitella teleta*, the cephalopod *Euprymna scolopes* and the platyhelminths *Schistosoma mansoni* and *Girardia tigrina,* all contain at least one Antp/Hox7-like and one Abd-A-like protein, but no *Drosophila* Ubx-like proteins.

In summary, two central Hox protein types are specific to protostomes: Abd-A type proteins can be found across all protostome clades while the presence of sequences of the *Drosophila* Ubx type is limited to the arthropod lineage.

**The central Hox protein type Antp/Hox7 is common to both protostomes and deuterostomes**

Only a single central Hox protein type (Antp/Hox7) could be identified as being present across both protostome and deuterostome lineages. With the exception of the *Drosophila* Ubx type, which is most similar to the Abd-A type proteins found in protostomes, all other central Hox protein types (vertebrate Hox6 & Hox8, echinoderm/hemichordate Echi/Hemi7 & Echi/Hemi8 and protostome Abd-A) are consistently more similar to Antp/Hox7-like proteins than to any other of the central Hox protein groups.

As expected from what is known about Hox proteins, no Hox proteins resembling Dfd/Hox4, Scr/Hox5 or the above central Hox sequence types were identifiable in species outside the bilaterian clade.

**Discussion**

The data presented here have numerous implications regarding the evolution of central Hox proteins and comparing their roles in establishing the central regions of the bilaterian animal body plans. The work of Malicki et al. [25] has often been cited as showing that Antp and Hox6 proteins are functionally equivalent, thereby providing support for the synteny-based classification and the assumption that the last common ancestor of protostome and deuterostome organisms possessed three central Hox proteins. This referencing is based on the similarity in phenotypes induced by ectopic expression of the corresponding proteins (generic leg induction and bristle pattern in T1). However, the authors of said manuscript worded their conclusions more carefully and specifically pointed out that the type of leg induced by Hox6 is not the same type of leg induced by Antp. Subsequent experiments performed by Percival-Smith et al. [26] showed that induction of generic legs is a feature common to most ectopically expressed Hox proteins. The second phenotype, the ability of other Hox proteins to induce the above mentioned bristle pattern in T1, has never systematically been examined. With this in mind, the claim that Antp and Hox6 proteins are functionally equivalent is not supported by experimental data, as this conclusion was based on their sharing an ectopic expression phenotype that is either now known to be common to most Hox proteins or was never systematically compared to other Hox proteins. Therefore, the core questions of how many central-type Hox proteins the last common ancestor of protostomes and deuterostomes possessed, when the various types of central Hox proteins arose and differentiated as well as which of the competing classification schemes best reflect the functional similarities across Hox proteins remain largely unresolved.

The work presented in this manuscript provides the first large-scale sequence-comparison-based analysis of the Hox protein family that is able to resolve and differentiate between the various types of central Hox proteins across protostomes and deuterostomes. Previous work by Thomas-Chollier et al. [27], aimed at classification of vertebrate Hox proteins, was able to differentiate between the central Hox proteins in the vertebrate lineage. Their later work [28], extending their analysis to encompass sequences from both protostomes and deuterostomes, however no longer resolved the vertebrate central Hox proteins, instead, grouping them into one large Hox6/7/8 group of paralogs. It is noteworthy to remark that, although using a very different approach, the latter work of Thomas-Chollier also identified the Antp/ftz/Lox5 proteins in the protostome lineage as the most similar to the vertebrate central Hox proteins, but did not resolve which specific vertebrate central Hox was most similar to Antp/ftz/Lox5 proteins. By using an approach based on pairwise data and calculating pairwise sequence similarities over a region known to be relevant to Hox protein function, yet previously neglected from analysis due to the difficulty of generating multiple sequence alignments encompassing this region, we were able to further resolve the classification of central Hox proteins from species in the protostome as well as deuterostome lineages. The ability to differentiate between different types of Hox sequences and having available a rough estimate of the earliest timepoint at which the respective sequence types are likely to have arisen, is a crucial prerequisite to being able to use information about putative similarities in sequence and/or function of Hox proteins and differentiate between conserved, independently or convergently acquired features. The sequence similarity based groupings of anterior Hox proteins Hox1/Lab Hox2/Pb, Hox3, Hox4/Dfd, Hox5/Scr are consistent with the accepted phylogeny of the Hox proteins (and corresponding functional equivalence studies) and discrepancies between the phylogeny-, synteny- and sequence-similarity based classification are only apparent for the phylogenetically unresolved 'central' and 'abdominal' type proteins (supplementary Figure S3 depicts the conflict between synteny and sequence-similarity data used to classify the protostome and deuterostome central Hox proteins). The observations that all major protostome and deuterostome clades contain a central Hox protein of the Antp/Hox7 type and that nearly all clade specific central Hox protein types (Hox6, Hox8 Echi/Hemi7, Echi/Hemi8 and Abd-A) are most similar to sequences of the Antp/Hox7 type leads us to conclude that Antp/Hox7 type sequences best represent the central Hox protein sequence type present in the last common ancestor of the protostome and deuterostome lineage. The other six central Hox protein types identified are specific to clades branching after the protostome-deuterostome split and therefore represent more recently derived members of the central Hox group.

<u>Deuterostome lineage:</u> The most deeply branching point at which we can identify well differentiated Hox6 and Hox8 type sequences is within the vertebrate lineage after the splits that gave rise to the vertebrate, urochordate and cephalochordate lineages. As Hox7 type proteins can be found in all major deuterostome clades, Hox6 and Hox8 type proteins appear to have diverged further from the ancestral-type central Hox protein in sequence, and presumably in function, than Hox7. Similarly, the most deeply branching point at which we can identify well differentiated Echi/Hemi7 and Echi/Hemi8 sequences is within the echinoderm and hemichordate lineages after their split from the chordates, which also indicates a divergence in sequence and function of these proteins from the ancestral-type central Hox protein. More importantly, since all major deuterostome clades possess proteins of the Antp/Hox7 type and all other deuterostome clade specific central Hox protein types (Hox6, Hox8 Echi/Hemi7 and Echi/Hemi8) are more similar to sequences of the Antp/Hox7 type than to any other Hox protein type, the vertebrate Hox6 and Hox8 as well as the echinoderm/hemichordate Echi/Hemi7 and Echi/Hemi8 are likely to be derived versions of a duplicated/triplicated Antp/Hox7 type protein in an ancestor of the respective lineages.

<u>Protostome lineage:</u> The most deeply branching point at which we can identify well differentiated *Drosophila* Ubx type proteins is within the arthropod lineage. Even within the Ubx-type sequences, two distinct sequence-similarity groups are formed (the first including proteins of the Ubx-IA isoform (isoform E), which is specific to dipterans, and the second including proteins of the Ubx type-IVA (isoform B) found in insects, chelicerates and crustaceans). Balavoine et al. presents two evolutionary scenarios for Lox2, Lox4, Abd-A and Ubx proteins. In one of them, the annelid Lox2 would be an ortholog of *Drosophila* Ubx [10], yet as shown in Figure 3, Lox2 and Ubx cluster in different sequence-similarity groups based on the sequence region we analyze. That the linker region of Lox2 differs from arthropod Ubx type proteins and is more similar in sequence to arthropod Abd-A type proteins than to their Ubx-type proteins could be interpreted as fitting with Balavoine's second proposed evolutionary scenario, in which the Lox2, Lox4, Abd-A and Ubx proteins we observe today are derived from a single protein present in the last common ancestor of protostomes. The observation that Ubx is arthropod specific and that dipterans have a lineage specific isoform of the protein also supports the claim that Ubx is a rapidly evolving Hox protein (crustacean and insect Ubx proteins are known to exhibit different molecular functions) [29]. The observation that the Ubx isoforms (Ia and IVA) form distinct sequence-similarity groups also fits well with comparative studies showing that these isoforms have different functions in *Drosophila* [30, 31]. We cannot, however, rule out the existence of further Ubx type sequences in other protostomes as, for some lineages, sequences or even sequence fragments from only a single species were deposited in the public database. This is most notable for onychophoran and lophophorate sequences for which only the species *Acanthokara kaputensis* and *Lingula anatina* were present respectively (note: within the analyzed YPWM-linker-homeodomain region the deposited onychophoran sequence annotated as "Ubx" is more similar to Abd-A type than Ubx-type seqences). In contrast, sequences of the Antp/Hox7 and Abd-A central Hox types are present in nearly all protostome lineages including the ecdysozoan and lophotrochozoan clades, indicating that the last common protostome ancestor likely possessed at least two differentiated central type Hox proteins, one of the Antp/Hox7 and one of the Abd-A type. The Abd-A group of proteins we define based on sequence similarity, also encompasses sequences annotated Lox4. Lox2 proteins cannot be assigned to any specific sequence similarity group, but are more similar to Abd-A type sequences than arthropod Ubx-type seqences. The above protostome specific Abd-A+Lox4, Lox2 and Ubx grouping is compatible with the hypothesis by Balavoine et al 2002 [10] that the Ubd-A peptide containing group of Hox proteins may have arisen from a central-type Hox protein after the protostome/deuterostome split. Based on our data, this ancestral protostome specific protein would best be represented by proteins of the Abd-A group. However we cannot provide insights as to how or when the different UbdA-motif containing sequences Abd-A, Ubx, Lox2 or Lox4 might have arisen. In this analysis we derive our sequence similarities only from features present across all Hox proteins (i.e. YPWM-linker-homeodomain) and thus do not take into account lineage specific peptide elements, such as the UbdA-peptide, that could help further resolve these groups.

The ancestral central type Hox protein: The identification of different sequence similarity groups in both the deuterostome and protostome lineages and the correlation of these groupings with the taxonomic distribution of the species from which the sequences were sampled, makes it unlikely that the last common ancestor of protostomes and deuterostomes possessed an even partially differentiated triplet of central Hox proteins. With the exception of the Antp/Hox7 type sequences, all other central type Hox sequence similarity groups consistently combine only sequences from species that diverged after the split of protostomes and deuterostomes. It is striking to see that the one type of central Hox proteins that is present in all protostome and deuterostome lineages (Antp/Hox7) is also the one type to which all other central Hox proteins show the greatest similarity (with the single exception of the rapidly evolving Ubx proteins). This indicates that all these groups were derived from an ancestral protein nowadays best represented by sequences of the Antp/Hox7 type.

We propose two possible explanations for these observations: 1) First, the last common ancestor of protostomes and deuterostomes possessed only a single Antp/Hox7 type sequence and subsequent duplication and divergence of this protein gave rise to the lineage-specific forms. 2) Second, the last common ancestor of these lineages already possessed multiple copies of an Antp/Hox7 type protein that thereafter were subject to very different selective constraints and therefore evolved divergently to form the lineage-specific forms. In either case, functional comparisons of central Hox proteins between protostome and deuterostome species can only be expected to yield information about features that have remained conserved since the protostome-deuterostome split, making any of the lineage specific Hox protein types suboptimal choices for such comparisons.

The old and the new classifications: Previously, central Hox proteins from protostomes and deuterostomes were either classified as one phylogenetically unresolvable group or, alternatively, classified by synteny as three groups of orthologous and functionally equivalent sequences, i.e. the orthologous pairs being Hox6 & Antp, Hox7 & Ubx, Hox8 & Abd-A [7, 32–35] (see Figure 1). Based on the pairwise-sequence-similarity clustering, we could identify seven distinct sequence types for the central Hox proteins: one present in both protostomes and deuterostomes (Antp/Hox7), one present in all protostomes (Abd-A), one specific to arthropods (Ubx), two specific to vertebrates (Hox6, Hox8) and two previously undescribed types of Hox proteins specific to the echinoderm and hemichordate lineages (Echi/Hemi7, Echi/Hemi8) (see Figure 3). The sequence-similarity-based classification presented here corresponds well to the taxonomic distribution of the species from which the sequences originated (i.e. no paraphyletic groupings occurred). The most deeply branching point at which we can identify Hox protein sequences of the Dfd/Hox4, Scr/Hox5 and central Hox types is within the bilateria. This observation is consistent with the hypothesis that only bilaterians possess central and central-like Hox proteins [36–40]. Our identified sequence similarity groupings also consistently reflect monophyletic groupings in the species classification which indicates that the sequence-similarities we base our classification on are not due to random effects, but correlate with functional and/or evolutionary constraints that are likely to have affected the sequence-evolution of the protein family. However, it is necessary to note that this analysis represents a snapshot of the currently available data. Not all bilaterian clades are equally well represented in the public databases and, with the accumulation of more sequence data for the poorly sampled clades, it is possible that further lineage specific Hox protein types remain to be discovered.

**Conclusion**
The accuracy of the often depicted "synteny-based" classification scheme should be reevaluated in light of our findings. While the more anterior Hox proteins are indeed highly conserved in both homeodomain sequence and protein function, the analysis we present indicates that the central Hox proteins across protostomes and deuterostomes are more diverse than previously assumed [7, 32–35] and that the Hox7/Antp-like proteins provide the least derived form of the putative ancestral central Hox protein that gave rise to the different types of central Hox sequences we observe today.

The ability to differentiate between conserved and more recently evolved protein types can provide additional sets of synapomorphic traits to classify species by and the ability to accurately identify which proteins diverged in sequence and which retained ancestral sequence features is crucial to elucidating the link between protein sequence and protein function for the Hox proteins. To accurately represent all of the available information, a classification scheme for the family of Hox proteins, combining sequences from both protostomes and deuterostomes, should either mark central Hox proteins as being phylogenetically unresolved or, if the aim is to identify putative functionally similar proteins, depict a sequence-data-based classification resolving the central Hox proteins. In the latter case the classification should ideally be based on a sequence region known to be relevant to protein function and present across all Hox proteins being compared while still providing sufficient information to resolve the central Hox proteins, such as the classification based on the YPWM-linker-homeodomain region provided in this manuscript.

**Methods**
Retrieval of Hox proteins:
A flow-chart overview of the approach is depicted in supplementary Figure S4. To identify all central Hox protein sequences present in the NCBI-nr database (National Centre for Biotechnology Information non-redundant GenBank protein database, May 20th 2010, ftp.ncbi.nlm.nih-.gov/blast/db/FASTA/nr.gz), we used an iterated PSI-BLAST search (version 2.2.22) [41] (inclusion value -h $10^{-30}$, results returned up to e-values of 10). The eight *Drosophila melanogaster* Hox proteins were used as individual queries and searched against the NCBI-nr database and run to convergence (Lab: 5 iterations, Pb: 7 iterations, Dfd: 13 iterations, Scr: 11 iterations, Antp: 12 iterations, Ubx: 19 iterations, Abd-A: 16 iterations and Abd-B: 2 iterations). From these searches, all high-scoring segment pairs (HSPs) with e-values up to 10 to any of the eight *D. melanogaster* query sequences were combined to ensure the inclusion of all sequences that might be relevant to our analysis (50585 non-redundant sequences). The corresponding full-length sequences for all hits were extracted.

Aim was to start from a well-defined set of Hox proteins in the protostome lineage and see whether, and to which extent, we could identify the well-described sets of known Hox proteins from the vertebrate lineage, thereby giving us an estimate for how likely we were to miss other Hox protein sequences of putative interest. The above approach proved highly successful, as the search broadened to a point that we retrieved many homeodomain protein sequences from well outside the Hox protein family (NK, Paired/Pax, Wox, TALE, Lim, etc.) It is therefore unlikely that, using this approach, we missed any of the Hox protein family sequences present in the database.

Identifying YPWM, linker+homeodomain regions:
The sequence region that previously provided the highest resolution classification for the central type Hox proteins is the region containing the YPWM (or FPWM), linker and homeodomain (YPWM-linker-homeodomain) [11]. To extract this region from the full-length sequences, we derived a Profile-Hidden-Markov-Model (HMM) from an alignment of the YPWM-linker-homeodomains for the eight *D. melanogaster* Hox proteins (programs used: AlnEdit [42] and Muscle [43]). The alignment was manually curated and a global HMM was derived from this alignment using HMMer [44]. The resulting HMM was calibrated with 5000 replicates and used to identify the corresponding YPWM-linker-homeodomains in the Hox-related full-length sequences (13282 sequences provided hits to the HMM with e-values better than 10). All Hox proteins known in the major model organisms (Mus musculus, Danio rerio, Branchiostoma floridae, Caenorhabditis elegans, Drosophila melanogaster), could be recovered using this Drosophila-centric approach, indicating that our HMM was general enough to adequately identify the homeodomains of all Hox proteins. The very relaxed E-value cutoff (10) was chosen to minimize the chance of excluding false-negative sequences from our subsequent analyses. The NCBI-nr database contained a number of Hox protein concatamers that are unlikely to represent any Hox proteins present in nature, as Hox proteins contain only one homeodomain. We therefore removed sequences containing more than one homeodomain from the dataset (13049 sequences remained). The remaining sequences were subsequently analyzed using CLANS [21].

CLANS clustering (identification of Hox protein sequence similarity groups):
CLANS provides a visual representation, in 3D, of the pairwise similarities of all sequences to each other using a force-directed layout/clustering approach (Fruchterman-Reingold [45]). Sequences are represented as dots and the pairwise sequence similarities are visualized as lines connecting the respective dots. Higher similarities are represented by darker lines and correspond to higher pairwise attractive forces. The higher the pairwise similarity between two sequences, the closer these two sequences tend to be located in 3D space. Chance similarities have a negligible effect in such a large map as they are averaged out by the sheer number of pairwise similarities being analyzed. Only groups of sequences that exhibit a systematic degree of pairwise similarity across many of their sequences/members are pulled together into clusters. This facilitates the visual selection of groups of sequences with higher than average similarity to one another. As shown in Figure 2 (left side), most of the Hox protein sequences form well separated clusters that are easily identifiable (e.g. Hox1,Hox2,Ho3,Hox9-13). In comparison, Dfd, Scr and the central group Hox proteins are much more similar to one another than any of the other groups and therefore are treated as one large cluster in this view. However, when examining this cluster in greater detail (Figure 2, right side), Dfd+Hox4, Scr+Hox5, and the central Hox protein sequences readily resolve into well-separated cluster (in the 3D view and, for the core members of each cluster, also apparent under more stringent P-value cutoffs in the 2D views (Fig S7)). The sequence-based groupings produced by CLANS are inherently stable and robust and reproducible groupings are formed even when specific sequence groups or all sequences of a given taxonomic clade are removed (see supplementary Figure S5 and S6, respectively). Similarly, the relative location of the various sequence groups changes only little over a wide range of different p-value cutoffs (see supplementary Figure S7). The p-value cutoff chosen for the analysis ($P=1e^{-30}$) was the one providing a good visual separation of all sequence clusters, while excluding as little data as possible.

Comparison of clustering results and species phylogeny:
All sequences of Dfd+Hox4, Scr+Hox5 and central-like Hox proteins were examined in detail and the overwhelming majority could be assigned to one of the ten sequence-similarity clusters shown in Figure 2 (right side). Supplementary Figure S8 depicts profiles derived for each of the Hox sequence similarity clusters, thereby providing a visual representation of the sequence signatures that led to the respective clusters. A previously published phylogeny of the species [20] was used as a basis to map the presence of the respective central Hox protein types onto a cladogram depicting the major protostome and deuterostome lineages (Figure 3). Specifically, we marked the earliest branch-point in the cladogram at which subsequent lineages evidenced a specific type of Hox protein. Multiple representative species were analyzed for each of the protostome and deuterostome clades to avoid potential artefacts arising from sparse sampling. The relative position of Hox genes in their respective genomic Hox gene cluster(s) was determined by examining the available genomes for sequenced organisms of the various lineages (vertebrates, arthropods, cephalochordates and nematodes). Additional information was retrieved from the literature for cephalochordates [46], urochordates [47], echinoderms [48], arthropods [49] and annelids [50]. The major taxonomic groupings we used in our analysis correspond to the NCBI taxonomic clades: Chordata, Cephalochordata, Urochordata, Hemichordata, Echinodermata, Arthropoda, Onychophora, Nematoda, Priapulida, Chaetognatha, Annelida, Mollusca, Platyhelminthes and Lophophorata (see supplementary Figure S2). The complete set of species that were manually validated regarding their taxonomic assignment and presence of Hox protein types are available in the CLANS save-files (see link below). Figure 3 summarizes the above data by depicting which Hox protein types could be identified in which of the bilaterian lineages.

Data files:
The CLANS files on which this analysis was based, the version of CLANS used for the analysis and the alignments used in the supplementary Figures S3 and S8 are available for download from http://bioinformatics.uni-konstanz.de/HueberHox/centralHox/index.php (User: centralHox; Password: origin. Login and password are case sensitive. Login requirement will be removed prior to publication). A web-page providing a 3-D view of the location of the sequences in the CLANS map is available at http://bioinformatics.uni-konstanz.de/HueberHox/centralHox/viewer/index.html (requires a WebGL enabled browser e.g. Firefox (version 4 or above) or Google Chrome). Due to the amount of data involved and the limitations on memory, graphics and GPU/CPU when displaying data via a web-browser, only the positions of the sequences/dots are shown. For full viewing and analysis capabilities please use the CLANS program (http://bioinformatics.uni-konstanz.de/programs/clans/index.php).

**Competing interests**
no competing interests to declare

**Authors' contributions**

SDH conceived the study, carried out the sequence-similarity analysis, phylogenetic profiling and drafted the manuscript. MAD provided material and analysis tools, participated in the study design, coordination and manuscript preparation. JR participated in the analysis and provided materials and analysis tools. HG participated in the analysis and manuscript preparation. GFW provided material, participated in the study design and manuscript preparation. TF provided material and analysis tools, participated in the study design, coordination, analysis and manuscript preparation. All authors have read and approved the final manuscript.

**Acknowledgements**

Funding for this work has been provided by the Australian Research Council, Center for Excellence Grant (CEO348212) and the Universität Konstanz. The funders had no role in the study .

**Figures:**

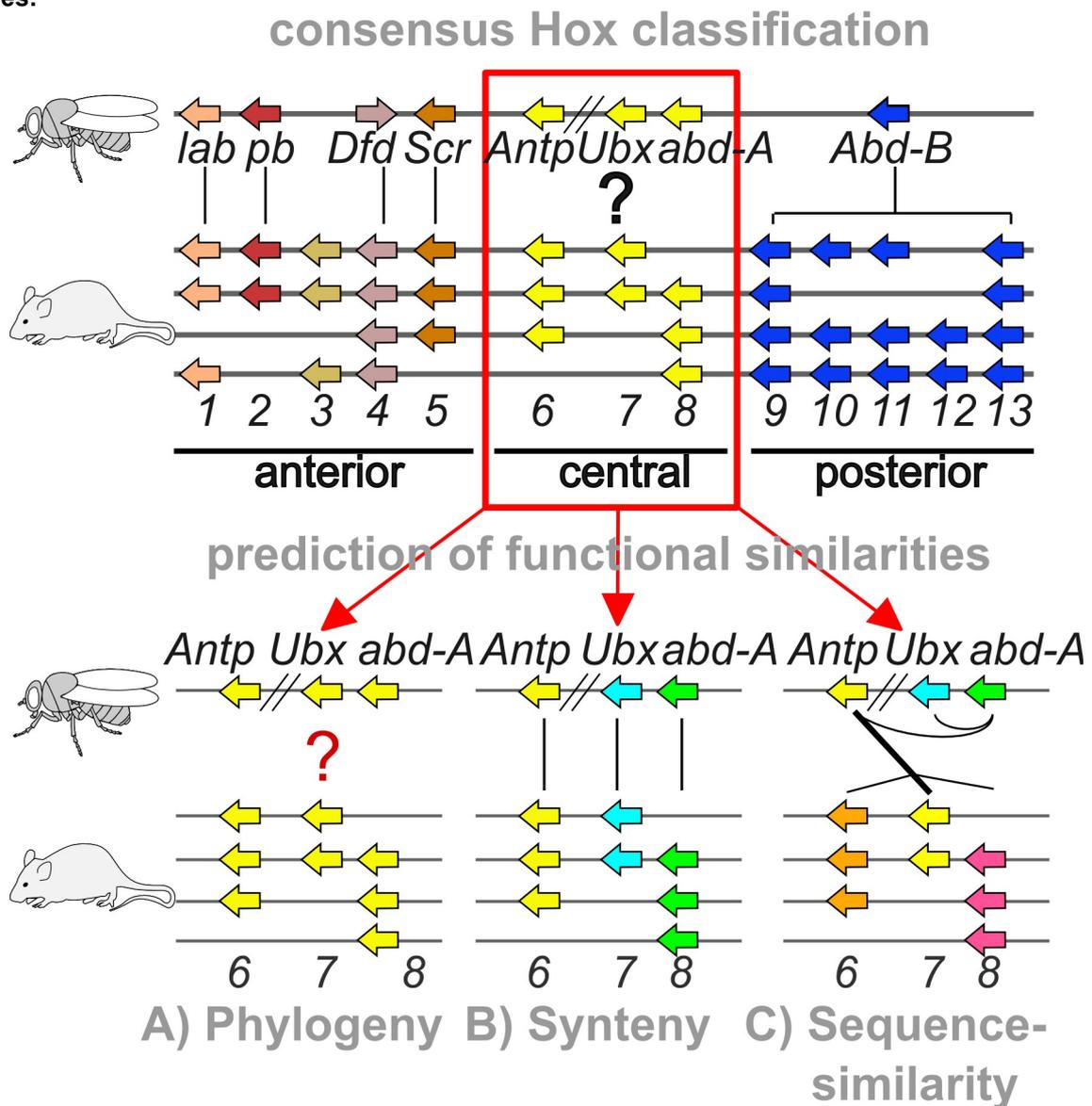

**Figure 1: Classification of central Hox genes.**

Top: Current consensus classification of the mouse *Mus musculus* and the fruit fly *D. melanogaster* Hox genes according to phylogenetic inference of the encoded protein sequence homeodomain. Relationships between the Antp, Ubx, Abd-A and Hox6, Hox7 and Hox8 proteins cannot be fully resolved. Bottom: Inferring functional similarity of the encoded proteins for the central Hox genes. A) Phylogeny does not fully resolve the relationships of the encoded central Hox proteins and therefore does not provide any statement regarding functional similarity of these proteins [7, 51]. B) The Synteny based classification postulates that the relative location of the Hox genes within the Hox cluster reflects their ancestry and function. It predicts a one-to-one orthology scenario with Antp being orthologous to Hox6, Ubx to Hox7 and Abd-A to Hox8 and, consequently, these protein pairs also as most similar in function [7, 33, 34]. C) Pairwise-sequence-similarity identifies the most sequence-similar proteins in the phylogenetically unresolved central group as *Drosophila spp.* Antp and vertebrate Hox7 proteins. The observed sequence-similarity pattern is compatible with a scenario assuming co-orthology of the proteins, but not with the postulated synteny classification. Based on the sequence similarities, the highest functional similarity across these proteins is predicted for Antp and Hox7 proteins [11].

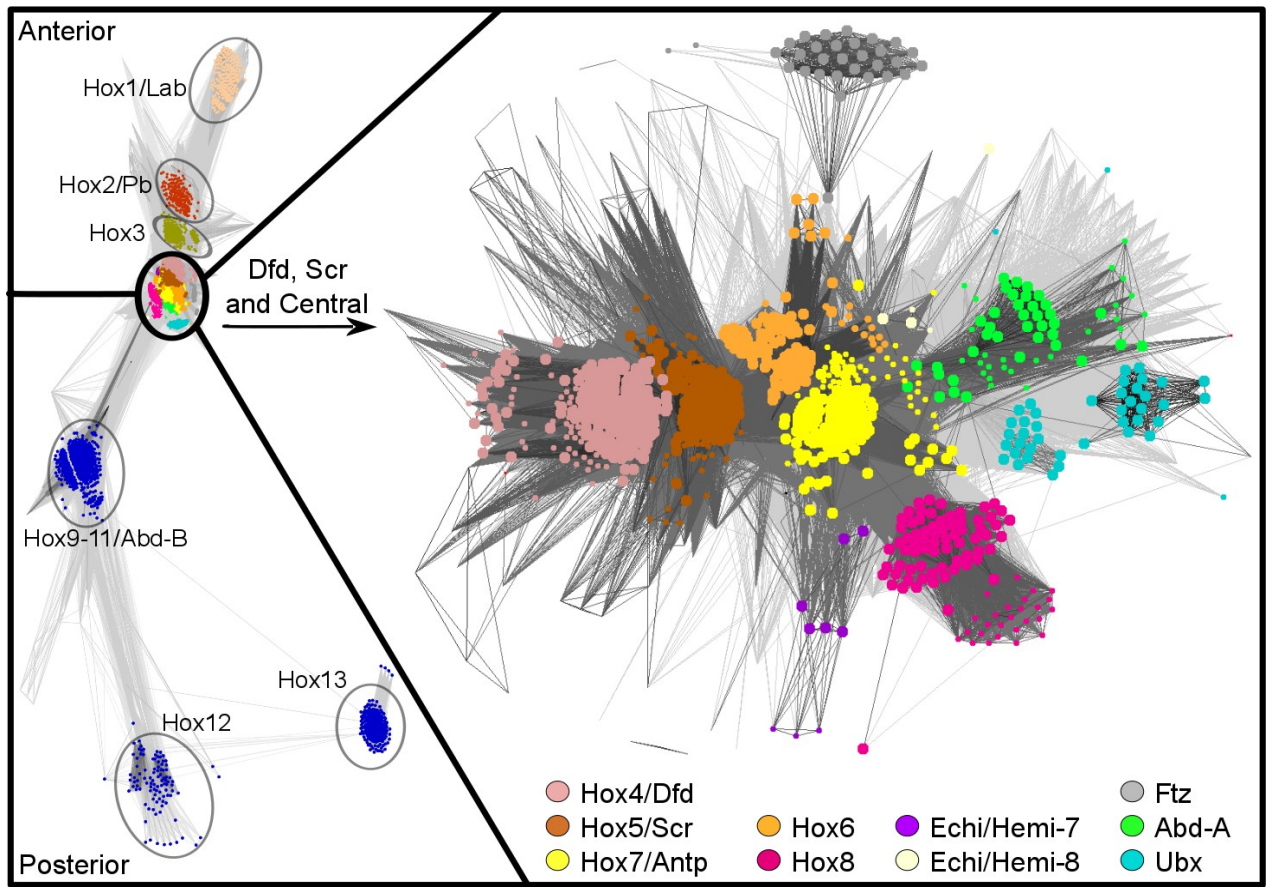

**Figure 2: 2D representation of the Hox protein sequence-similarities.**

CLANS representation of pairwise sequence-similarities for the 2629 sequences identified as belonging to the core Hox-group. Sequences are represented as dots and lines connecting the dots represent their pairwise similarities. The more similar the sequences, the darker the line, the stronger the attraction between dots and the more closely they are located to each other in the map. Left side: CLANS map derived for the core group of Hox and Hox-like proteins (2629 sequences, similarity p-value cutoff = $1e^{-23}$). Coloring of the dots is according to the classification scheme shown in Figure 1. Right side: CLANS map focusing on the Hox4, Hox5, Hox6, Hox7, Hox8, Dfd, Scr, Antp, Ftz, Abd-A, Ubx and sequence similar Hox proteins that formed one compact cluster in the map displayed on the left (1095 sequences, similarity p-value cutoff = $1e^{-30}$). Ten separate clusters are identifiable and color-coded according to the types of sequences present in each group. Large colored dots represent sequences containing the YPWM-motif, linker region and homeodomain, while small dots represent truncated or fragment sequences missing one or more of these elements. Dots without any coloring represent sequences that could not be unambiguously assigned to one specific sequence similarity group. The depicted maps are shown in 2D for technical reasons. The sequence groups were assigned based on a 3D-view of the dataset providing an additional discriminatory dimension. The 2D and the 3D versions of the sequence maps are provided in the supplementary materials, as is an overview of the similarities and differences between the vertebrate and arthropod sequence similarity groups (supplementary Figure S3).

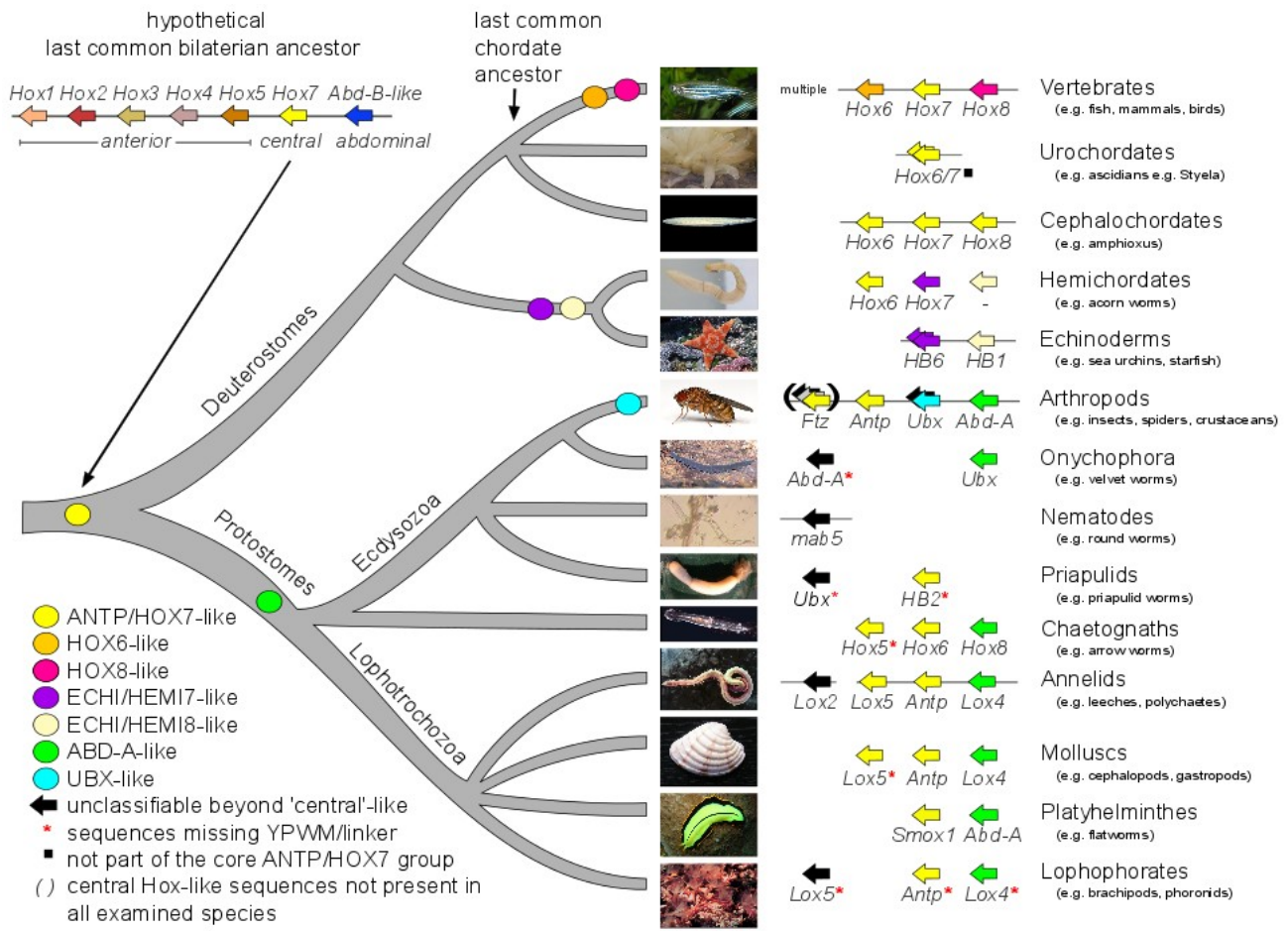

**Figure 3: Distribution of central Hox proteins across protostome and deuterostome lineages.**

The left side depicts a cladogram approximating the major taxonomic divisions in the protostome and deuterostome lineages, adapted from Dunn et al., 2008 [20] and the presumed first occurrence of the different central Hox protein types (circles, color-coded as in Figure 2). The right side depicts the genes (arrows), chromosomal arrangement and annotated names for the different types of central Hox proteins present in the various clades (consensus depicted) (sources for the pictures are listed in supplementary Table ST1). Horizontal lines connecting Hox genes indicate that these are represented according to their relative locations on the chromosome. Missing horizontal lines indicate an absence of data regarding the relative location of the genes in the genome. Positional information was determined by examining the genomic location of the Hox genes in sequenced organisms (vertebrates, arthropods, cephalochordates and nematodes). Additional positional information was received for cephalochordates [46], urochordates [47], echinoderms [48], arthropods [49] and annelids [50]. Identifiable isoforms (e.g. via FlyBase) are not shown. Sequences with a near identity to other sequences in the same species yet differing in more than three amino acids, i.e. potential splice variants or recently duplicated genes, are represented as slightly shifted arrows of the same color. No central-like Hox proteins could be identified in species outside the bilaterian clade.

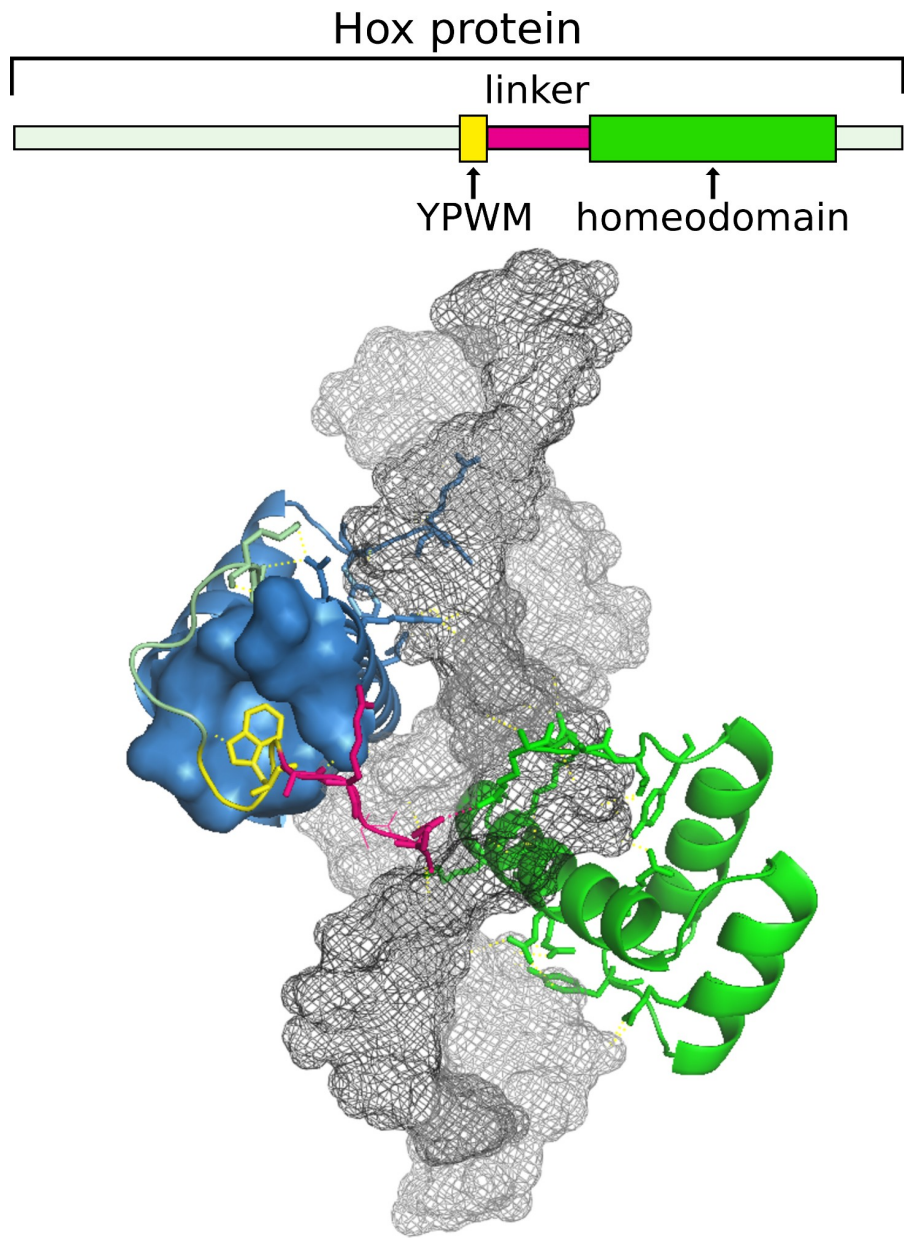

**Supplementary Figure S1: Molecular structure of Hox proteins.**

Schematic representation and structure of the Hox protein elements that determine the binding of Hox proteins to DNA, adapted from Hueber, 2009 [52] in PyMol [53]. These elements include the YPWM motif, the linker region and the homeodomain, shown here for the Hox protein Scr, its interaction partner Exd (a TALE class homeodomain protein) on the $fkh^{250}$ enhancer of the *forkhead* gene (structure 2R5Z [54]). The DNA is shown in "mesh" structure in dark and light grey. Both Scr and Exd are shown in cartoon structures, Exd in blue and differing colors for the Scr domains: The YPWM motif is coloured in yellow, the linker region in pink and the homeodomain in green. The yellow YPWM motif interacts with a hydrophobic pocket of Exd (surface model), the homeodomain (green) interacts with the major and minor groove of the DNA, while the linker region (pink) interacts with both the protein Exd and the minor groove of the DNA. Amino acids in the linker region interacting with either the DNA or Exd are shown in "stick" structures. A section of the linker region as well as the homeodomain of the Scr protein are not shown, as no structural information was available for the corresponding residues.

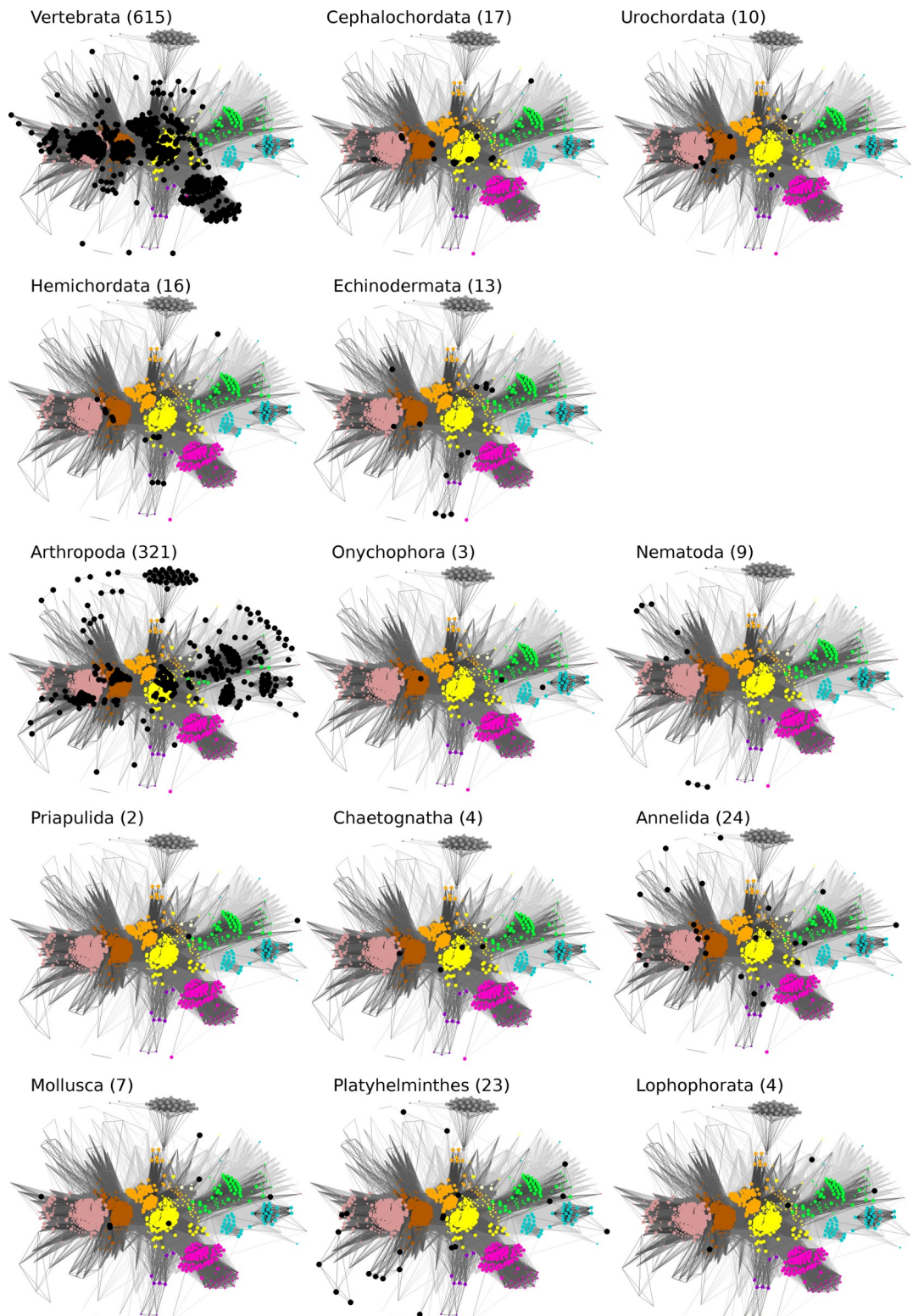

**Supplementary Figure S2: Distribution of the central Hox protein sequence types across the bilaterian clades.**

2D pairwise sequence similarity clustering of the Hox4-8, Dfd, Scr, Antp, Abd-A, Ubx-type and similar sequences. The coloring scheme is the same as used in Figure 2. Black circles highlight sequences of the species corresponding to the NCBI taxonomic clade stated above each tiling (corresponding number of circles is shown in brackets).

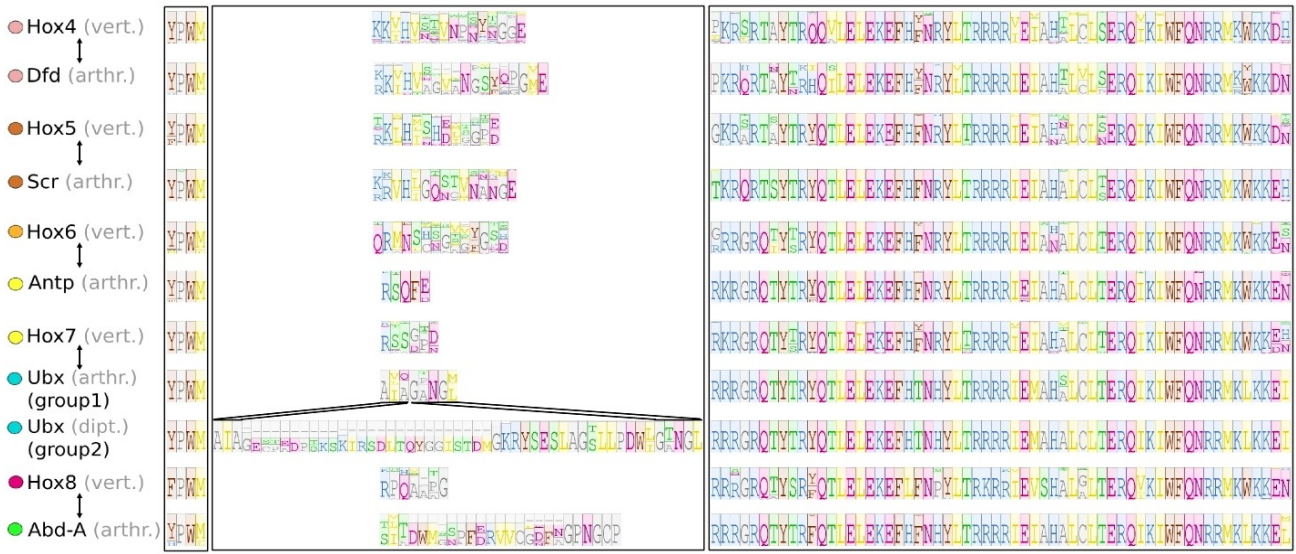

**Supplementary Figure S3: profiles of arthropod and vertebrate central Hox proteins.**

Sequence profiles for the central Hox protein groups (Hox6, Hox7, Hox8, Antp, Abd-A and Ubx) are depicted and ordered in this figure according to the synteny scheme to highlight the discrepancy between the synteny-based groupings (vertical bidirectional arrows) and the sequence-similarity based groupings (colored dots). Positively charged amino acids are colored in blue, negatively charged in pink, hydrophobics in yellow, aromatics in brown and potentially phosphorylated amino acids in green. All others are colored in grey. The multiple sequence alignments for each group were created via AlnEdit [42] and visualized using MotifDraw [55]. Alignment regions identified as 'inserts', e.g. in which at least 50% of the sequences had gaps, are not shown as some sequences contained large inserts either within or outside the homeodomain.

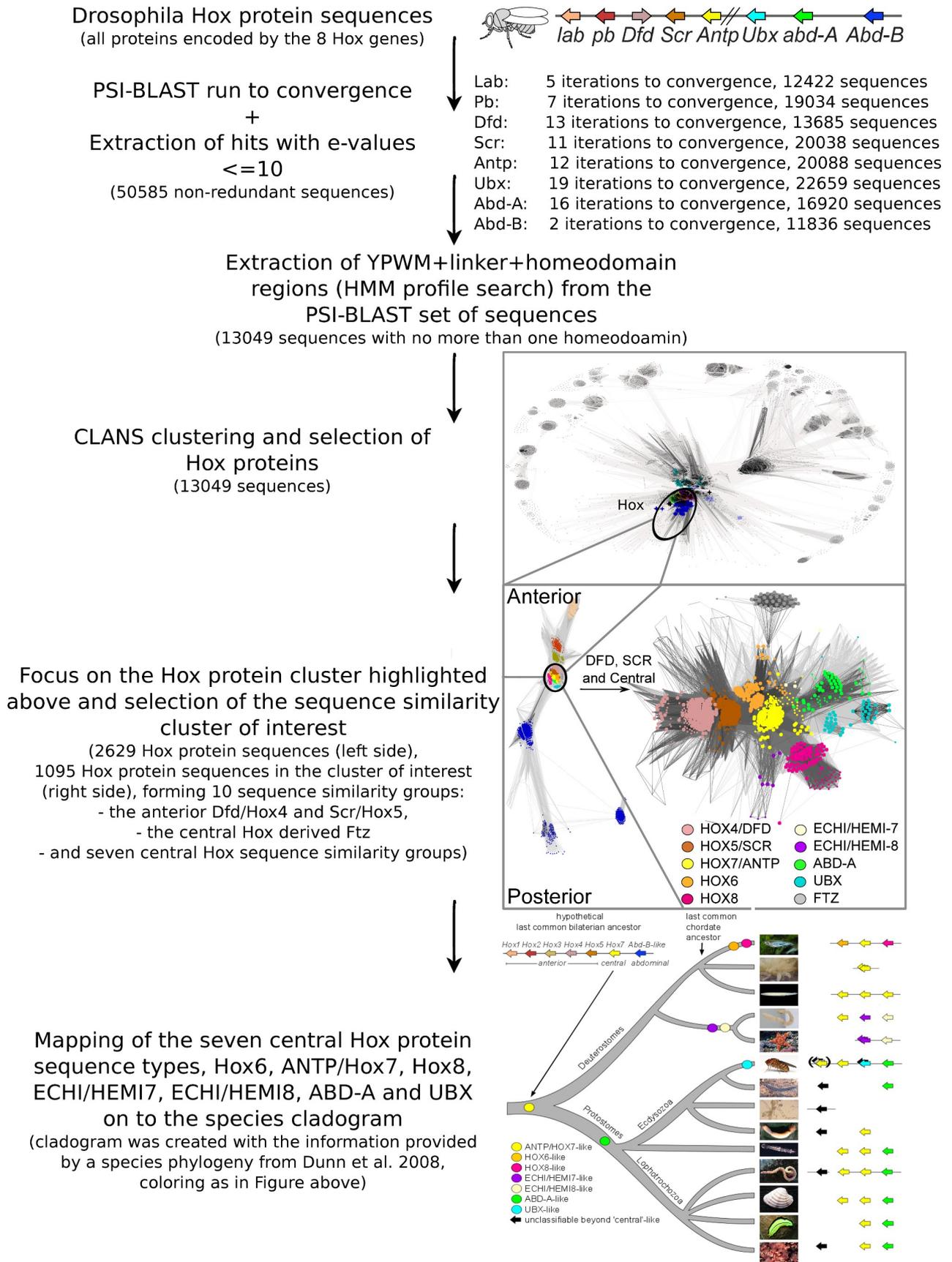

**Supplementary Figure S4**: Flow-chart overview of the CLANS classification approach for central type Hox proteins.

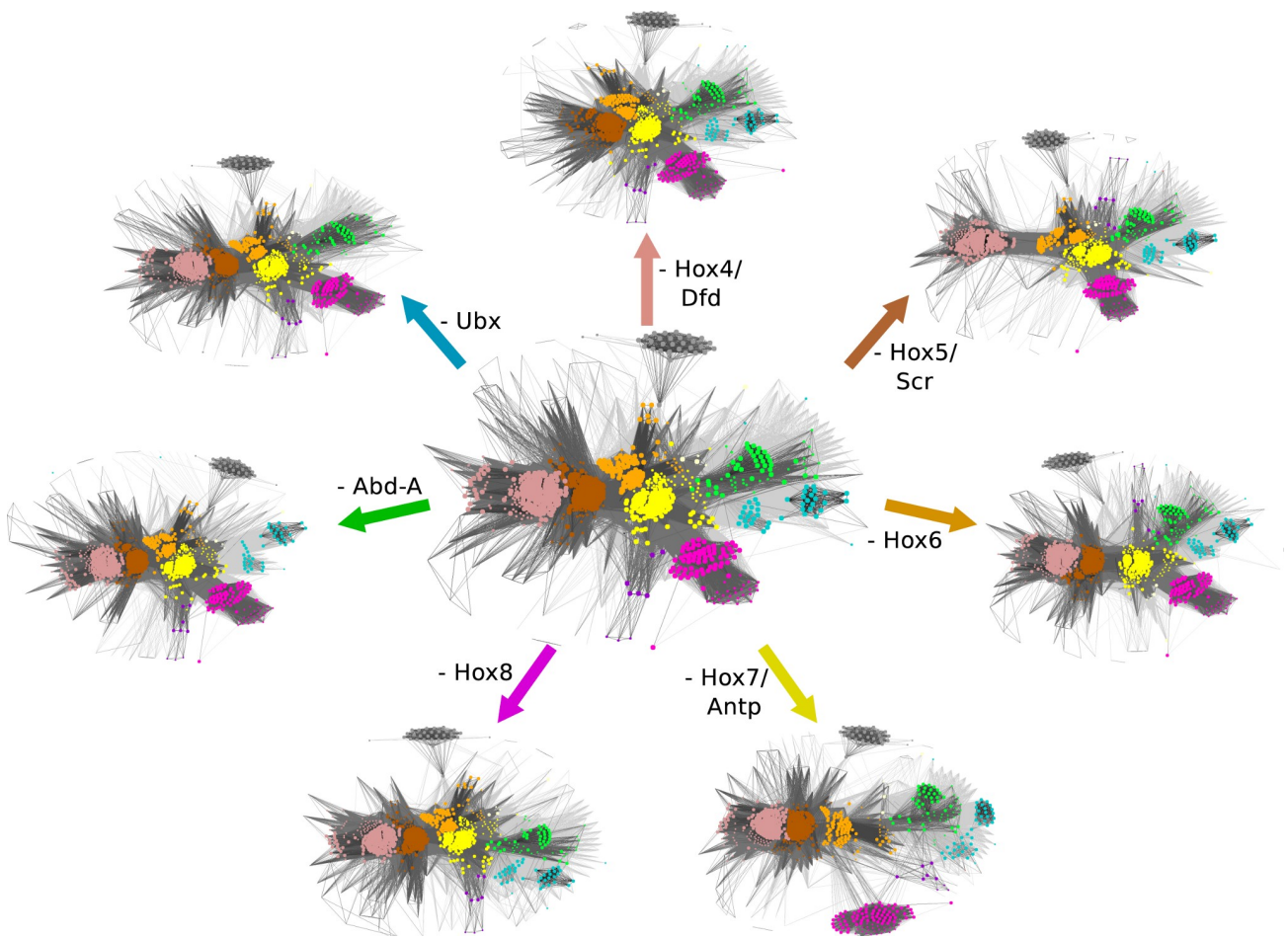

**Supplementary Figure S5: Stability/reproducibility of CLANS maps after removal of individual sequence groups.**

The central graph depicts the clustering of the complete set of sequences of interest. The graphs at the periphery display the sequence similarity clusters generated after individual sequence similarity groups were removed from the dataset. The Figure above demonstrates that the sequence clusters formed are inherently stable, meaning that even after removal of all sequences of a given sequence type, the groupings of the remaining sequences are largely unaffected. Each CLANS picture shown above was independently initialized (the sequence dots were randomly placed in a 2D space) and run until no further change in the clustering was detectable. The complete graphs were then rotated/mirrored to place Ftz proteins (grey) at the top and Hox4/Dfd (light-red) to the left. The sequence similarity p-value cutoff used in all graphs was $1e^{-30}$.

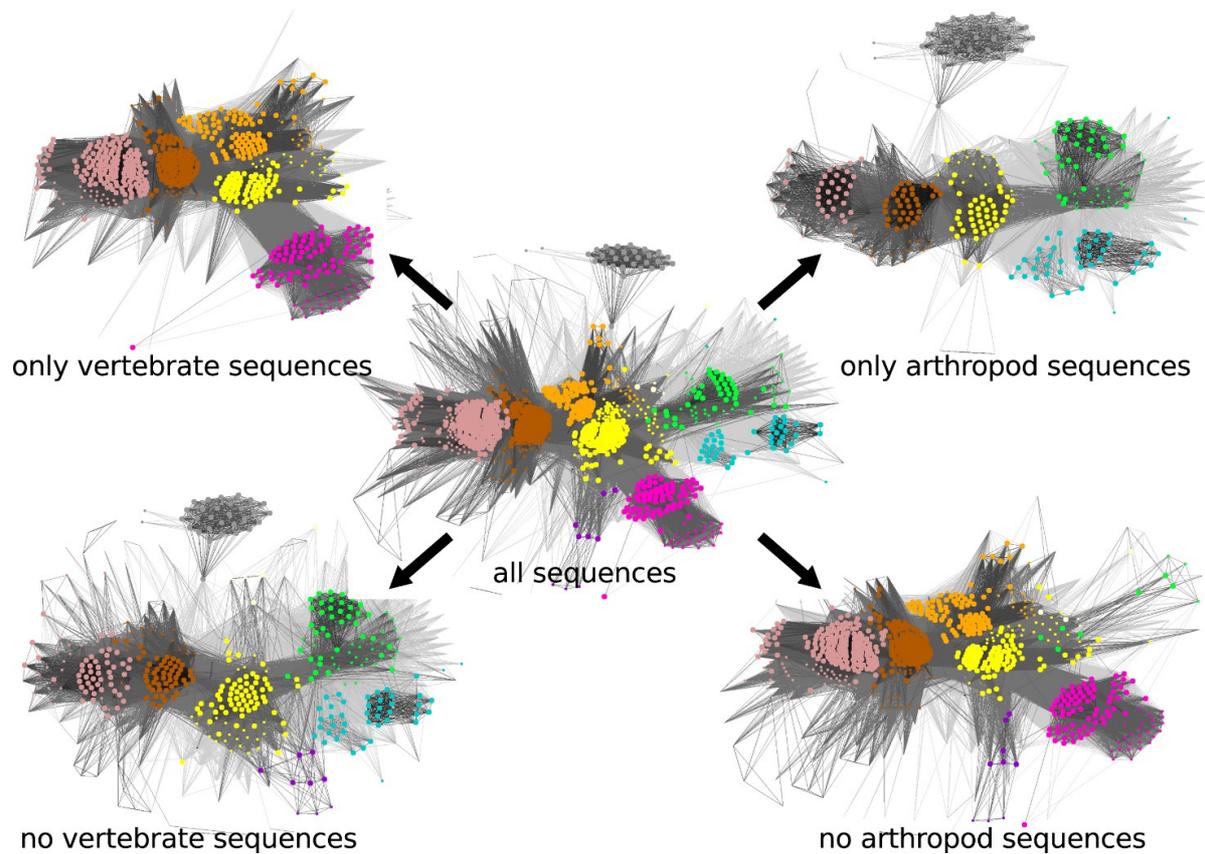

**Supplementary Figure S6: Stability/reproducibility of CLANS approaches after removal of sequences of specific taxonomic clades.**

The Figure depicts the clusters formed by Dfd/Hox4, Scr/Hox5 and the central Hox proteins sequences in absence of vertebrates or arthropod sequences as well as those formed by only vertebrate or only arthropod sequences. The figure above demonstrates that even after removal of all sequences of specific taxonomic clades the overall grouping of the remaining sequences remains unaffected. Each CLANS picture shown above was independently initialized (the sequence dots were randomly placed in 2D space) and run until no changes in the clustering were detectable. The complete graphs were then rotated/mirrored to place Ftz proteins (grey) at the top and Hox4/Dfd (light-red) to the left. The sequence similarity p-value cutoff used in all graphs was $1e^{-30}$.

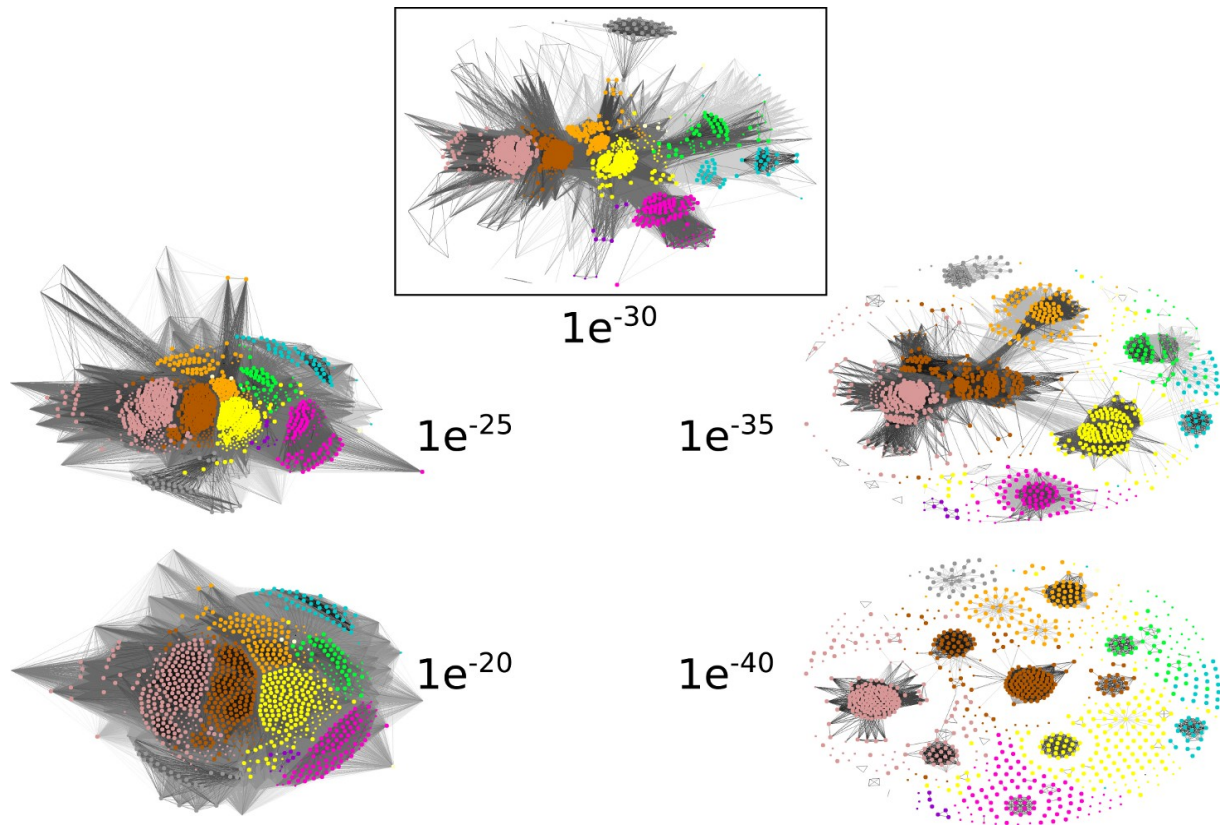

**Supplementary Figure S7: Stability/reproducibility of CLANS approaches with changing p-value cutoffs.**

CLANS graphs of the Hox4-8, Dfd, Scr, Antp, Abd-A and Ubx type sequences (central and central-like) for a range of different p-value cutoffs ($1e^{-20}$ to $1e^{-40}$ in $1e^{-5}$ steps). The relative relationships between sequence groups remain the same over a wide range of different p-values. However, at p-values more permissive than $1e^{-30}$ (the cutoff we used for our analysis), the sequences crowd together and separate groups are not easily detectable ($1e^{-25}$ and $1e^{-20}$). At p-values more stringent than $1e^{-30}$, many sequences lose all connections to other sequences and the remaining sequence groupings are strongly reduced in size ($1e^{-35}$ and $1e^{-40}$). At $1e^{-35}$ the clear separation of the core members of the Hox6 vs Hox7/Antp type sequence groups in more easily apparent.

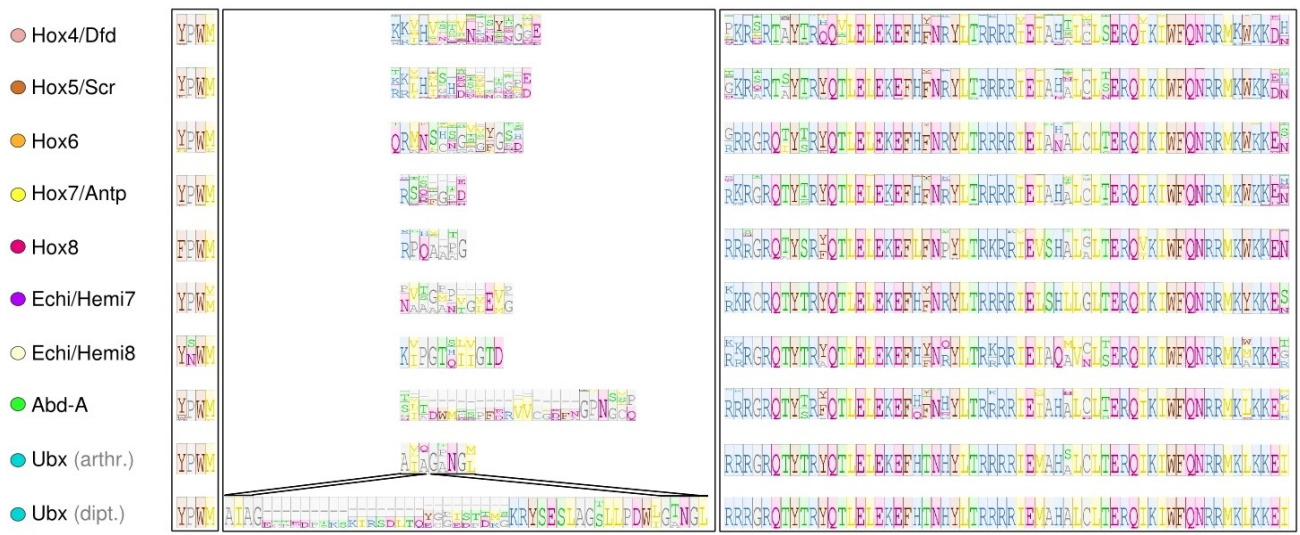

**Supplementary Figure S8: profiles for the Hox4/Dfd, Hox5/Scr and central Hox protein clusters.**

Sequence profiles were generated for each of the identified groups to provide an overview of the differences and similarities across the sequences contained within the individual sequence similarity groups (Figure 2). Only sequences containing the complete YPWM, linker + homeodomain region were used to generate the above sequence profiles. Positively charged amino acids are depicted in blue, negatively charged in pink, hydrophobics in yellow, aromatics in brown and potentially phosphorylated amino acids in green. All other residues are colored in grey. The multiple sequence alignments for the respective groups were created using AlnEdit [42] and visualized using MotifDraw [55]. Alignment regions identified as 'inserts', e.g. in which at least 50% of the sequences had gaps, are not shown as some sequences contained large inserts within and outside the homeodomain.

**Supplementary Table ST1**

| Picture for | Species | Source | License & Author |
|---|---|---|---|
| Vertebrates | *Danio rerio* | http://en.wikipedia.org/wiki/File:Zebrafisch.jpg | Copyrighted free use {Author: Azul} |
| Cephalochordates | *Branchiostoma lanceolatum* | Personal communication, Michael Schubert, Institut de Génomique Fonctionnelle de Lyon, France | Copyrighted free use {Author: Michael Schubert} |
| Urochordates | *Ciona intestinalis* | http://en.wikipedia.org/wiki/File:Cionaintestinalis.jpg | GFDL / Creative Commons Attributions-Share Alike 3.0 Unported {Author: Perezoso} |
| Hemichordates | *Saccoglossus spec.* | http://en.wikipedia.org/wiki/File:Eichelwurm.jpg | GFDL / Creative Commons Attribution-Share alike 3.0 Unported {Author: Necrophorus} |
| Echinoderms | Asteroidea | http://en.wikipedia.org/wiki/File:Nerr0878.jpg | Public Domain / no restrictions {Author: NOAA} |
| Arthropods | *Drosophila repleta* | http://en.wikipedia.org/wiki/File:Drosophila_repleta_lateral.jpg | GFDL / Creative Commons Attributions-Share Alike 3.0 Unported {Author:Bbski} |
| Onychophora | *Peripatoides indigo* (?) | http://en.wikipedia.org/wiki/File:Onycophora_(515525252).jpg | Creative Commons Attribution-Share Alike 2.0 Generic {Author: Bruno Vellutini} |
| Nematodes | nematode | http://en.wikipedia.org/wiki/File:Roundworm.jpg | Public Domain / no restrictions {Author: Josh Grosse} |
| Priapulids | *Priapulus caudatus* | http://en.wikipedia.org/wiki/File:Priapulus_caudatus.jpg | Creative Commons Attribution 3.0 Unported {Author: Dmitry Aristov} |
| Chaetognaths | *Spadella cephaloptera* | http://en.wikipedia.org/wiki/File:Chaetoblack.png | Creative Commons Attribution 3.0 Unported {Author: Zatelmar} |
| Annelids | *Glycera sp.* | http://en.wikipedia.org/wiki/File:Nerr0328.jpg | Public Domain / no restrictions {Author: NOAA} |
| Molluscs | *Chione paphia* | http://commons.wikimedia.org/wiki/File:Bivalvia.jpg | Creative Commons Attribution-Share Alike 3.0 Unported {Author:u|Bricktop} |
| Platyhelminthes | *Pseudoceros dimidiatus* | http://en.wikipedia.org/wiki/File:Pseudoceros_dimidiatus.jpg | Creative Commons Attribution-Share Alike 2.0 Generic {Author: Richard Ling} |
| Lophophorates | phoronid | http://en.wikipedia.org/wiki/File:Nur03506.jpg | Public Domain / no restrictions {Author: NOAA} |